\documentclass[]{aa}
\usepackage{natbib}         
\usepackage{graphicx}
\usepackage{longtable}
\usepackage{xcolor}
\bibpunct{(}{)}{;}{a}{}{,} 

\def\Qlim{25}\def\nombreq{5200}

\newcommand{\ind}[1]{_{\mathrm{#1}}}
\newcommand{\diff}{\mathrm{d}}

\newcommand\num{\nu\ind{m}}
\newcommand\nup{\nu\ind{p}}
\newcommand\nug{\nu\ind{g}}
\newcommand\Dnu{\Delta\nu}\newcommand\Dnup{\Delta\nu\ind{p}}
\newcommand\Tg{\Delta\Pi_1}

\newcommand\Teff{T\ind{eff}}

\newcommand\mpm{\mathcal{M}}
\newcommand\XR{X\ind{R}}

\newcommand\dnurot{\delta\nu\ind{rot}}

\newcommand{\eva}{{\mathcal{E}}}
\newcommand{\dens}{{\mathcal{N}}}

\newcommand\numax{\nu\ind{max}}

\newcommand\Pm{\tau}

\newcommand\thetap{\theta\ind{p}}
\newcommand\thetagg{\theta\ind{g}}
\newcommand\np{{n\ind{p}}}

\newcommand{\BV}{Brunt-V\"ais\"al\"a}
\newcommand{\NBV}{N\ind{BV}}

\newcommand{\dq}{\delta q}

\newcommand{\rin}{r_1}
\newcommand{\rout}{r_2}

\newcommand\Kepler{\emph{Kepler}}

\newcommand{\modif}[1]{\textbf{#1}} 
\renewcommand{\modif}[1]{\textcolor{blue}{#1}}

\begin{document}

\title{Period spacings in red giants}
\subtitle{III. Coupling factors of mixed modes}
\titlerunning{Coupling factors of mixed mode}
\authorrunning{Mosser et al.}
\author{%
 B. Mosser\inst{1},
  C. Pin\c con\inst{1}, K. Belkacem\inst{1}, M. Takata\inst{2},
  M. Vrard\inst{3}
 }
 \institute{
 LESIA, Observatoire de Paris, PSL Research University, CNRS, Universit\'e Pierre et Marie Curie,
 Universit\'e Paris Diderot,  92195 Meudon, France
 cedex, France; \texttt{benoit.mosser@obspm.fr}
 \and
 Department of Astronomy, School of Science, The University of Tokyo, 7-3-1 Hongo, Bunkyo-ku, Tokyo 113-0033, Japan
 \and
 Instituto de Astrof\'isica e Ci\^encias do Espa\c co, Universidade do Porto, CAUP, Rua das Estrelas, 4150-762 Porto, Portugal
}


\abstract{The power of asteroseismology relies on the capability
of global oscillations to infer the stellar structure. For evolved
stars, we benefit from unique information directly carried out by
mixed modes that probe their radiative cores. This third article
of the series devoted to mixed modes in red giants focuses on
their coupling factors that remained largely unexploited up to
now.}
{With the measurement of the coupling factors, we intend to give
physical constraints on the regions surrounding the radiative core
and the hydrogen-burning shell of subgiants and red giants.}
{A new method for measuring the coupling factor of mixed modes is
 set up. It is derived from the method recently implemented for
measuring period spacings. It runs in an automated way so that it
can be applied to a large sample of stars. }
{Coupling factors of mixed modes were measured for thousands of
red giants. They show specific variation with mass and
evolutionary stage. Weak coupling is observed for the most evolved
stars on the red giant branch only; large coupling factors are
measured at the transition between subgiants and red giants, as
well as in the red clump.}
{The measurement of coupling factors in dipole mixed modes
provides a new insight into the inner interior structure of
evolved stars. While the large frequency separation and the
asymptotic period spacings probe  the envelope and the core,
respectively, the coupling factor is directly sensitive to the
intermediate region in between and helps determining its extent.
Observationally, the determination of the coupling factor is a
prior to precise fits of the mixed-mode pattern, and can now be
used to address further properties of the mixed-mode pattern, as
the signature of the buoyancy glitches and the core rotation.}


\maketitle

\section{Introduction}

The seismic observations of large sets of stars with the CoRoT and
\Kepler\ missions, from the main sequence
\citep{2011Sci...332..213C} up to the red giant branch
\citep{2009Natur.459..398D}, has motivated intense work in stellar
physics, among which is ensemble asteroseismology
\citep[e.g.][]{2010A&A...522A...1K,2010A&A...517A..22M,2011ApJ...743..143H,2011ApJ...740L...2S,2014A&A...570A..41K}.
Ensemble asteroseismology is efficient for evolved stars as they
host mixed modes that behave as gravity modes in the core and as
pressure modes in the envelope. These modes directly probe the
stellar core and, therefore, reveal unique information.

Contrary to pressure modes, evenly spaced in frequency, and to
gravity modes, evenly spaced in period, mixed modes show a more
complicated frequency pattern. Since, for red giants, the density
of gravity modes is large compared to the density of pressure
modes, the mixed-mode pattern resembles a pure gravity-mode
pattern perturbed by the pressure-mode pattern. The period
spacings are close to the asymptotic value for gravity-dominated
mixed modes, but are significantly smaller near expected pure
pressure modes \citep[e.g.,][]{2015A&A...584A..50M}.
Pressure-dominated mixed modes have lower inertia than
gravity-dominated mixed modes, hence show larger amplitudes
\citep{2009A&A...506...57D,2014A&A...572A..11G}. Even with period
spacings far from the asymptotic values, they allowed us in a
first step to distinguish stars with hydrogen-burning shell from
stars with core helium-burning
\citep{2011Natur.471..608B,2011A&A...532A..86M}. In a second step,
the asymptotic analysis of the mixed-mode pattern allowed us to
derive precise information on the \BV\ frequency profile in the
radiative core \citep{2012A&A...540A.143M}. Indeed, the asymptotic
expansion
\citep{1979PASJ...31...87S,1989nos..book.....U,2012A&A...540A.143M,2016PASJ...68..109T}
is a powerful tool for investigating mixed modes in red giants
observed by the space missions CoRoT and \Kepler, despite the fact
that observations are not conducted in an asymptotic regime.
Observed radial pressure orders are small, too small for lying in
the asymptotic regime, as shown by CoRoT observations
\citep[e.g.,][]{2009Natur.459..398D,2010A&A...517A..22M}. However,
\cite{2013A&A...550A.126M} have demonstrated that the second-order
asymptotic expansion provides a coherent view on the pressure mode
pattern, even for the most evolved red giants
\citep{2013A&A...559A.137M}. Conversely, high radial gravity
orders are observed in mixed modes, except in subgiants, so that
considering a first-order expansion for the gravity contribution
is relevant \citep{2014A&A...572L...5M}.

This work allowed us to measure asymptotic period spacings and to
derive unique information on stellar evolution. The contraction of
the helium core of hydrogen-shell-burning red giants is marked by
the decrease of the asymptotic period spacing. Moreover, stars
with a degenerate helium core (with a mass inferior to about
1.8\,$M_\odot$) on the red giant branch (RGB) show a close
relationship between the large separation and the period spacing,
which is the seismic signature of the mirror effect between the
core and the envelope structures. In the red clump, core-helium
burning stars with a mass lower than about 1.8\,$M_\odot$ show a
tight mass-dependent relation between the asymptotic large
separation and period spacing, contrary to more massive stars.

The use of the mixed modes for assessing the inner interior
structure of red giants is, however, still in its infancy. Up to
now, only the period spacings have benefitted from large scale
measurements \citep{2016A&A...588A..87V}. Another important
parameter, the coupling factor $q$ was precisely determined for a
handful of stars only \citep{2016A&A...588A..82B}. This parameter
benefitted from a recent breakthrough since
\cite{2016PASJ...68..109T} could derive an asymptotic expression
for dipole modes using the JWKB method (Jeffreys, Wentzel,
Kramers, and Brillouin), taking the perturbation of the
gravitational potential into account, but without using the
weak-coupling approximation, contrary to the previous work by
\cite{1989nos..book.....U}.

Here, we use the method developed by \cite{2015A&A...584A..50M}
(Paper I of the series) in order to complete the large-scale
analysis of \cite{2016A&A...588A..87V} (Paper II of the series).
We specifically address the measurement of the coupling factor $q$
of mixed modes. This parameter plays an important role in the
asymptotic expansion: it expresses the link between the pressure
and gravity contributions to the mixed mode, respectively
expressed by their phases $\theta\ind{p}$ and $\theta\ind{g}$.
Following \cite{1989nos..book.....U}, the asymptotic expansion
reads
\begin{equation}\label{eqt-asymp}
    \tan\theta\ind{p} = q \tan\theta\ind{g}
    ,
\end{equation}
with the phases defined as
\begin{eqnarray}
  \thetagg &=& \pi {1 \over \Tg}  \left({\displaystyle{1\over\nu}
  -\displaystyle{1\over\nug}}\right), \label{eqt-g} \\
  \thetap &=& \pi {\nu-\nup\over \Dnup} \label{eqt-p},
\end{eqnarray}
where $\nup$ and  $\nug$ are the asymptotic frequencies of pure
pressure and gravity modes, respectively, and $\Dnup$ is the
frequency difference between the consecutive pure pressure radial
modes with radial orders $\np$ and $\np+1$, as defined by
\cite{2015A&A...584A..50M}.

While $\thetap$ and $\thetagg$ respectively account for the
propagation of the wave in the envelope and in the core, the
coupling factor comes from the contribution of the region between
the \BV\ cavity and the Lamb profile $S_\ell = \sqrt{\ell
(\ell+1)}\, c/r$ where the oscillation is evanescent. Hence,
studying the coupling factor $q$ provides a direct way to examine
the evanescent region, namely the physical regions surrounding the
hydrogen-burning shell, where most of the stellar luminosity is
produced. This parameter plays also a crucial role for examining
rotational splittings
\citep{2013A&A...549A..75G,2015A&A...580A..96D} and mode
visibilities \citep{lowV}, so that its thorough examination
becomes crucial in red giant seismology.

The article is organized as follows. Section \ref{formalism}
presents the framework of our analysis. We use the weak-coupling
cases in order to illustrate and emphasize how $q$ is linked to
stellar interior properties. We also introduce the strong-coupling
case, which is necessary at various evolutionary stages. In
Section \ref{method}, we develop a specific method for measuring
the coupling factors in an automated way. Section
\ref{observations} presents the results derived from the set of
\Kepler\ red giants analyzed by \cite{2016A&A...588A..87V}.
Results are discussed in Section \ref{discussion}.

\section{Coupling factor of mixed modes\label{formalism}}

In this Section, we show how the coupling factor of mixed modes
can be used to probe the stellar interior. We restrain the
analysis to dipole mixed modes ($\ell=1$).

\subsection{Evanescent region\label{evanescent}}

\begin{figure}
 \includegraphics[width=8.8cm]{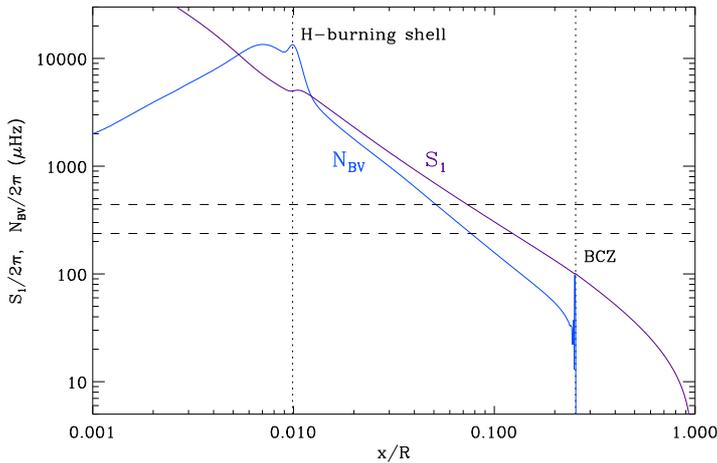}
 \caption{\BV\ and Lamb frequency profiles for a 1.3-$M_\odot$ stellar model on
 the RGB, corresponding to model M1 of \cite{2015A&A...579A..31B} (see also
 Table \ref{tab-modeles}). The horizontal dashed lines delimit the frequency range where oscillations
 are expected, around $\numax$; the vertical dotted lines indicate the locations of
 the hydrogen-burning shell and of the base of the convection zone
 (BCZ), respectively.}\label{fig-model}
\end{figure}

The coupling factor in Eq.~(\ref{eqt-asymp}) measures the decay of
the wave in the region between the cavities delimited by the \BV\
frequency $\NBV$ and  the Lamb frequency $S_1$
(Fig.~\ref{fig-model}). This intermediate zone is called the
evanescent region. In the JWKB approximation, thus assuming that
the wavelength of the oscillations is much shorter than the scale
height of the equilibrium stellar structure, the radial component
of the wavevector $\kappa$ in the evanescent region can be written
\begin{equation}\label{eqt-kappa}
    \kappa = {\sqrt{(S_1^2-\omega^2)(\omega^2-\NBV^2)}
    \over
    c \,\omega}
    ,
\end{equation}
where $c$ is the sound speed and $\omega$ is the angular
frequency.

We assume in the following that the evanescent zone is located
between the hydrogen-burning shell and the base of the convective
envelope. The \BV\ and Lamb frequencies show similar radial
variations in this region probed by the mixed modes
\citep[Fig.~\ref{fig-model} and, e.g., Fig. 2
of][]{2012ASSP...26...23M}. Both can be approximated with a power
law with similar exponent:
\begin{equation}\label{eqt-def-beta}
   - {\diff\ln \NBV\over \diff \ln r}
 = \beta_{N}
 \simeq \beta
 \simeq \beta_{S}
 = - {\diff\ln S_1\over \diff \ln r}
 .
\end{equation}
In fact, the exponent $\beta$ measures the density contrast
between the core and the envelope: the higher the contrast, the
higher $\beta$. We expect $\beta$ to increase with the evolution
and contraction of the core. In evolved red giants, where the
density contrast between the envelope and the core is high enough
to ensure that the local gravity varies as $r^{-2}$ in the
evanescent region, $\beta$ is expected to be close to the upper
limit of 3/2 \citep{2016PASJ...68..109T}.

As a consequence of the parallel variations of $\NBV$ and $S_1$,
the ratio
\begin{equation}\label{eqt-def-alpha}
    \alpha = \NBV / S_1
\end{equation}
can be considered as nearly uniform in the region above the
hydrogen-burning shell where the oscillation is evanescent. A very
thin evanescent zone means $\alpha$ close to unity, whereas a wide
one has a small $\alpha$. In the following, we use this ratio
$\alpha$ as a measure of the extent of the evanescent region and
aim at linking $\alpha$ and $\beta$ with $q$.

\subsection{Weak coupling\label{section-indep}}

In the case of weak coupling, the decay of the wave amplitude in
the evanescent region $\eva$ is expressed by the transmission
factor defined as
\begin{equation}\label{eqt-def-T}
  T \equiv \exp \left(-\int_\eva \kappa \, \diff r \right)
  ,
\end{equation}
and linked to the coupling factor $q$ of the asymptotic expansion
(Eq.~\ref{eqt-asymp}) by
\begin{equation}\label{eqt-couplage-Unno}
   q = {T^2\over 4}
   ,
\end{equation}
as computed by \cite{1979PASJ...31...87S} and
\cite{1989nos..book.....U}.

Weak coupling is ensured if the transition region between $S_1$
and $\NBV$ is wide enough. Indeed, as shown by
Eqs.~(\ref{eqt-def-T}) and (\ref{eqt-couplage-Unno}), the value of
$q$ in the case of weak coupling is necessarily below 1/4.
Conversely, a value of $q$ above 1/4 implies that the
weak-coupling approach is insufficient.

\begin{figure}
 \includegraphics[width=8.8cm]{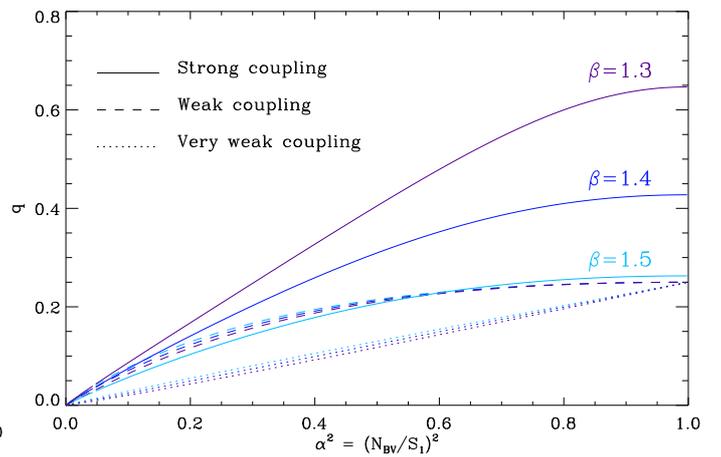}
 \caption{Relationship between the coupling factor $q$ and the ratio
 $\alpha^2 =(\NBV / S_1)^2$ defined under the assumption that both frequencies have the same slope
varying as $r^{-\beta}$ in the region surrounding the
hydrogen-burning shell; $q$ is computed for three different values
of $\beta$ (1.3, 1.4 and 1.5) in three cases: very weak (dotted
lines, Eqs.~\ref{eqt-approx-boundaries}-\ref{eqt-approx}), weak
(dashed lines, Eqs.~\ref{eqt-def-T}-\ref{eqt-indep}), or strong
coupling (full lines,
Eqs.~\ref{eqt-relation-T-q}-\ref{eqt-definition-X}). The very weak
and weak coupling values hardly depend on $\beta$, unlike the
strong-coupling case.}\label{fig-qNS}
\end{figure}

With the assumptions made in Section \ref{evanescent}, so with the
definitions of $\alpha$ (Eq.~\ref{eqt-def-alpha}) and $\beta$
(Eq.~\ref{eqt-def-beta}), the integral term in
Eq.~(\ref{eqt-def-T}) can be rewritten
\begin{equation}\label{eqt-indep}
 \int_\eva \kappa\ \diff r =  {\sqrt{2}\over  \beta} \int_\alpha^1
 \sqrt{1-x^2}\sqrt{x^2 - \alpha^2} \; {\diff x \over x^2}
 ,
\end{equation}
where $x$ is the normalized frequency $\omega / S_1$.
Consequently, the coupling factor depends on $\alpha$ and $\beta$
only, but does not vary in the frequency range where modes are
observed. Although the evanescent region is probed at different
depths by the different mixed modes, the wave transmissions
through the barrier, hence $q$, are the same for each frequency
because of the parallel variations of $\NBV$ and $S_1$. This
result confirms that $q$ is directly related to the interior
structure properties (Fig.~\ref{fig-qNS}).

In the case of very weak coupling, the influence of the turning
points can be neglected in Eq.~(\ref{eqt-def-T}). In practice, we
consider that $S_1 \gg \omega \gg \NBV$ almost everywhere between
the boundaries of the evanescent region $\rin$ and $\rout$, so
that $q$ reduces to a simple function of these boundaries
\begin{equation}\label{eqt-approx-boundaries}
    q \simeq {1\over 4} \left({\rin \over \rout}
    \right)^{2 \sqrt{2}}
    .
\end{equation}
Combined with the assumption on the variations of $\NBV$ and
$S_1$, it can be rewritten in terms of the coefficients $\alpha$
and $\beta$:
\begin{equation}\label{eqt-approx}
    q \simeq {1\over 4}\ \alpha^{2 \sqrt{2} / \beta}
    .
\end{equation}
This simplified relation shows again that $q$ does not vary in
frequency range where mixed modes are observed. It also proves
that $q$ provides another diagnostic parameter of red giants,
complementary to the period spacing $\Tg$ that probes the \BV\
cavity.

\subsection{Strong coupling\label{section-strong}}

In the strong coupling case studied by \cite{2016PASJ...68..109T},
the expression of the transmission $T$ is not as simple as
Eq.~(\ref{eqt-def-T}) and must be replaced by a more precise
expression. The relation between $q$ and $T$ becomes
\begin{equation}\label{eqt-relation-T-q}
    T^2 = {4 q\over (1+q)^2}
    .
\end{equation}
The factor $q$ is connected to interior structure properties in
the general case
\begin{equation}\label{eqt-relation-q-X}
    q = {1 - \sqrt{1- \exp ( -2\pi X)}
       \over
        1 + \sqrt{1- \exp (-2\pi X)}}
        ,
\end{equation}
following  Eq.~(133) of \cite{2016PASJ...68..109T}. The new
variable $X$ defined by Eq.~(61) of this paper expresses
\begin{equation}\label{eqt-definition-X}
    X \propto \int_\eva \kappa\,\diff r + \XR
    ,
\end{equation}
where the additional term $\XR$ is important only when the
frequencies $\NBV$ and $S_1$ are very close to each other. It
comes from the gradient of $\NBV$ and $S_1$ in the evanescent
region. In practice, it takes the reflection of the wave at the
boundaries into account and so explains that the transmission
cannot be equal to unity even when $\NBV$ and $S_1$ have very
close values.  As a result, a very narrow evanescent region
characterized by $\alpha\simeq 1$ is not associated with a
coupling factor close to one.

The variation of $X$ and the relation between $X$ and $q$ make it
possible to have values of $q$ significantly above the limit of
1/4 fixed by the weak-coupling case (Eq.~\ref{eqt-couplage-Unno}).
Under the assumption that $\NBV$ and $S_1$ show similar radial
variations, the computation of $X$, hence $q$, depends on $\alpha$
and $\beta$ only. The variation of $q$ with $\alpha^2$ are shown
in Fig.~\ref{fig-qNS}, computed with Eq.~(A80) of
\cite{2016PASJ...68..109T}. This equation takes into account the
perturbation of the gravitational potential that modifies the
values of $\NBV$ and $S_1$. The comparison of the weak and strong
coupling also shows that the use of the weak-coupling case is
relevant for very small values of $q$ only.

We note that, the lower the $\beta$, the larger the correction on
$q$ in the strong-coupling case. Moreover, the observations of $q$
values larger than 1/4 are associated to $\beta$ values less than
1.5. We also note that the correction increases when $\alpha$
increases.

As in the weak-coupling case, $q$ does not show variation in the
frequency range where modes are observed. This is the consequence
of parallel variations of $\NBV$ and $S_1$. In the following, we
consider that the variation of $q$ with frequency is small enough,
so that fitting the whole mixed-mode spectrum with a fixed
coupling factor makes sense.

\section{Method\label{method}}

In previous work \citep{2012A&A...540A.143M,2014A&A...572L...5M},
coupling factors in red giants were derived from the fit of the
oscillation pattern. This fitting method, even if made precise and
easy with the new view exposed in \cite{2015A&A...584A..50M},
cannot be automated, so that we have to provide a new method for
dealing with the amount of \Kepler\ data.

\subsection{Correlations between mixed-mode parameters}

The method of \cite{2016A&A...588A..87V} developed for measuring
the period spacing $\Tg$ offers an efficient basis for obtaining a
relevant measure of $q$. According to Eq.~(\ref{eqt-asymp}), the
measurements of the mixed-mode parameters $q$ and $\Tg$ are a
priori independent: $\Tg$ measures the period spacings between the
mixed modes whereas $q$ measures the deformation of these spacings
close to the pure pressure modes \citep[see, e.g., Fig. 2
of][]{2016A&A...588A..87V}. When buoyancy glitches, rotational
splittings, or simply noise,  modify locally the frequency
interval between two consecutive mixed modes, this independence is
however not ensured. In fact, they introduce a crosstalk between
the determination of $q$ and $\Tg$, which induces spurious
fluctuations of $q$ when computing $\Tg$.
\cite{2016A&A...588A..87V}, who considered $q$ as a free parameter
when measuring period spacings $\Tg$, obtained $q$ with large
uncertainties. In order to get $q$ in a robust manner, we had
first to circumvent this crosstalk. To do so, we chose to measure
$q$ and $\Tg$ in an independent manner. For measuring $q$, we
considered first $\Tg$ as a fixed parameter, adopting the values
of \cite{2016A&A...588A..87V}.

\begin{figure}
\includegraphics[width=8.8cm]{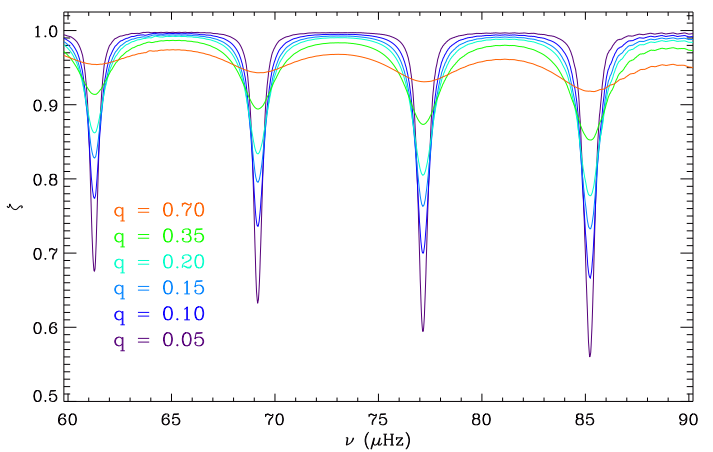}
\caption{Variation of the stretching function $\zeta$ with $q$.
The locations of the local minima of $\zeta$, which correspond to
the expected pure pressure modes, do not depend on the value of
$q$. } \label{fig-vari-q}
\end{figure}

\begin{figure}
\includegraphics[width=8.8cm]{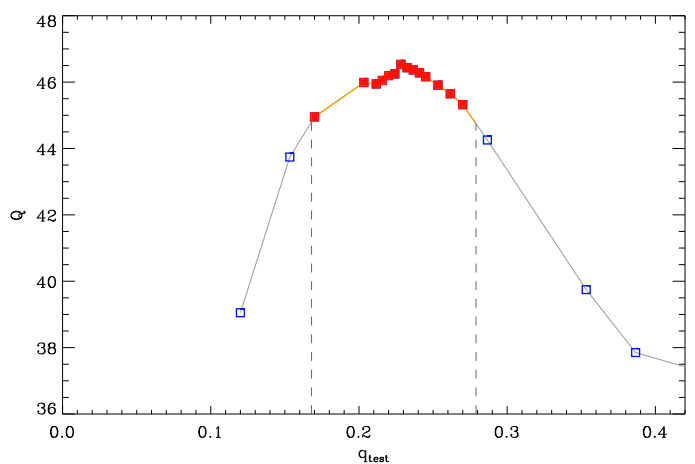}
\caption{Variation of the signal $Q$ with the coupling factor
$q\ind{test}$ used in the stretching function $\zeta$ considered
as a free parameter, for a typical clump star (KIC 1717994). The
region with maximum $Q$ values (full symbols) is used to define
the optimum value of $q$ and the associated uncertainty $\delta
q$. The dotted lines delimit the range $q\pm \delta
q$.}\label{fig-exemple}
\end{figure}

\subsection{Measuring $q$}

In practice, the first step of the method consists in a stretching
of the power spectrum $P (\nu)$ by using the change of variable
introduced by \cite{2015A&A...584A..50M}. The frequency $\nu$ is
replaced by the stretched period $\tau$ according to
\begin{equation}\label{eqt-stretch}
    \diff\Pm = {1\over \zeta} {\diff \nu \over \nu^2}
    ,
\end{equation}
where the function $\zeta$ is obtained from the interpolation of
the values obtained for the dipole mixed-mode frequencies $\num$,
\begin{equation}\label{eqt-zeta}
    \zeta(\num) = \left[1+ {1\over q}
     {\num^2 \Tg \over \Dnup}
    {\cos^2 \thetagg
    \over
    \cos^2 \thetap}
    \right]^{-1}
    .
\end{equation}
The way to obtain a precise continuous function of $\zeta$ is
explained by \cite{2015A&A...584A..50M}. It importantly depends on
the correct location of the pure dipole modes, carried out by the
use of the red giant universal pattern
\citep{2011A&A...525L...9M,2012ApJ...757..190C}.

The power spectrum $P (\tau)$ expressed as a function of the
stretched period $\tau$ is composed of $\mpm$ blended comb-like
patterns, where $\mpm$ is the number of visible azimuthal orders.
For a star seen equator-on, $\mpm=2$, only 1 when the star is seen
pole-on, and 3 in the intermediate case.

The regularity of the $\mpm$ comb(s) is optimized when the value
of $\zeta$ used for stretching the spectrum matches the correct
coupling (Fig.~\ref{fig-vari-q}). So, we varied the coupling
factor $q\ind{test}$ to obtain a varying correction $\zeta$ and
searched to optimize the regularity of the comb. The optimum
signal is inferred from the maximum of the Fourier transform of $P
(\tau)$, noted $Q$ (Fig.~\ref{fig-exemple}). As is clear from
Fig.~\ref{fig-vari-q}, the regularization is only slightly
affected by $q$, so that the measurement is possible for high
signal-to-noise spectra only.

\begin{figure*}
\includegraphics[width=16cm]{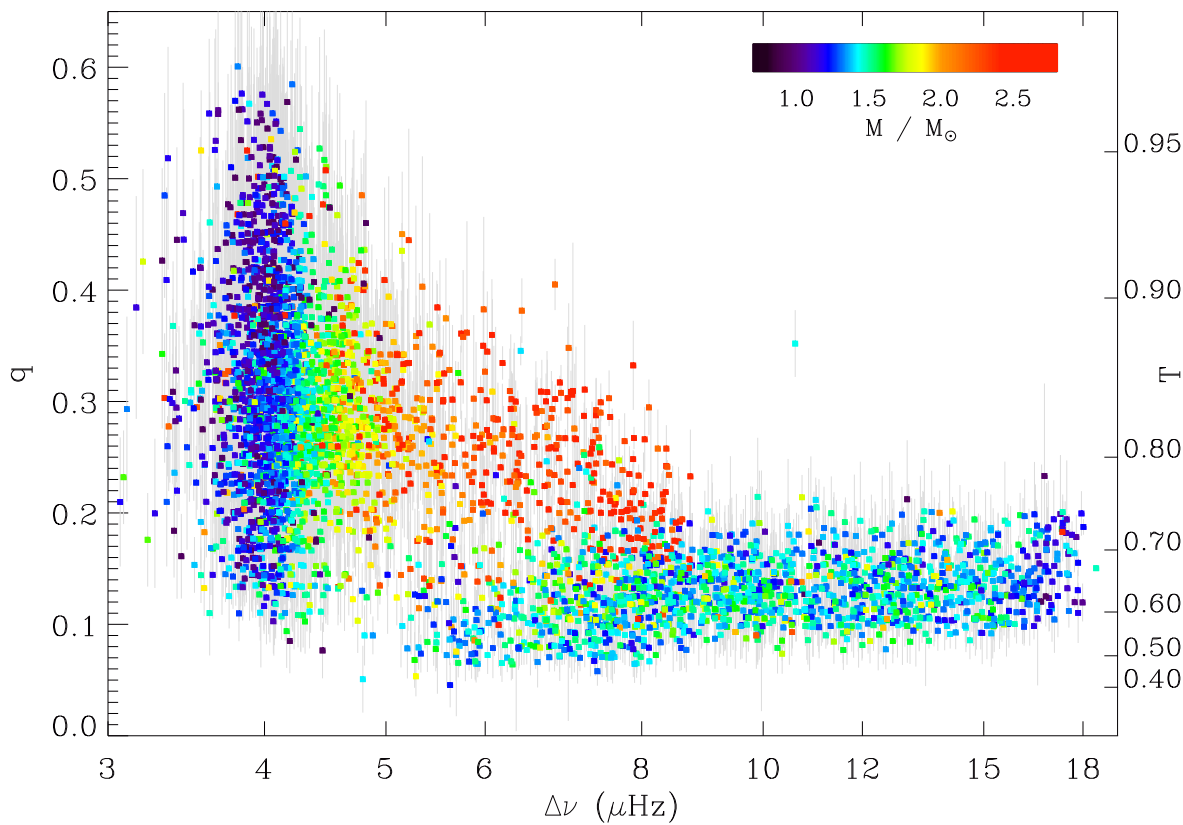}
\caption{Coupling factor $q$ as a function of the large separation
$\Dnu$. Asymptotic values of $\Dnu$ and $\Tg$ come from
\cite{2016A&A...588A..87V}. The color codes the mass, determined
with seismic scaling relations. The right vertical axis provides
the corresponding values of the transmission factor
$T$.}\label{fig-coupling}
\end{figure*}

\subsection{Individual check and limitations}

The robustness of the method was verified with individual checks
based on typical red giant oscillation spectra observed at various
evolutionary stages. This check first allowed us to verify that
more than 96\,\% of the prior values of $\Tg$ automatically
measured by \cite{2016A&A...588A..87V} are safe. Wrong initial
values of $\Tg$ were identified by spurious measures of $q$. For
oscillation spectra with a low signal-to-noise ratio, measuring
the period spacing $\Tg$ still remains possible, \modif{but
identifying the small variations due to the coupling is
demanding}. For high signal-to-noise ratio oscillation spectra, we
identified three major cases providing us with incorrect
measurements for $q$: pressure glitches, buoyancy glitches, and
rotation. All these effects perturb the regularity of the function
$\zeta$.

\subsubsection{Pressure glitches}

Pressure glitches occur when rapid variations of the sound speed
modify the regularity of pressure modes. They contribute by adding
a small modulation to the pressure mode pattern. The shift remains
less than 2\,\% of $\Dnu$, as measured in the radial modes of a
large set of red giants by \cite{2015A&A...579A..84V}. The
modulation of the pure dipole pressure modes is similar to that of
radial modes, so that the location of pressure-dominated mixed
modes are shifted. As a result, the local minima of the stretching
function $\zeta$ are shifted \citep[Fig. 7
of][]{2015A&A...584A..50M}. This change may induce variation of
$Q$ larger than the variation due to $q$, so that the method
fails. More importantly, the method often produced very low or
very large spurious values of $q$ in that case.

\subsubsection{Buoyancy glitches}

Buoyancy glitches occur when rapid variations of the \BV\
frequency modify the regularity of the gravity-mode pattern. This
effect was theoretically investigated by
\cite{2015ApJ...805..127C}. It affects mixed modes since it
modifies locally the period of the stretched spectrum, with a
clear signature on the function $\zeta$ \citep[Figs. 8 and 10
of][]{2015A&A...584A..50M}. Measuring a mean period spacing in the
presence of buoyancy glitches is often possible, but measuring its
optimization for deriving $q$ is more challenging, since
variations of the period spacings due to the buoyancy glitch mimic
variations due to an inadequate value of $q$. This situation most
often occurs for red clump stars and may explain part of the large
spread of $q$ observed for these stars.

\subsubsection{Rotational splittings}

Rotational splittings also perturb the measurements of $q$. In the
case of low rotation \citep{2012A&A...548A..10M}, each family of
stretched peaks associated to a given azimuthal order $m$ provides
a comb spectrum with a period close to $\Tg$ \citep[Fig. 6
of][]{2015A&A...584A..50M}, so that the identification of $\Tg$
and of $q$ derives clearly from the optimization of the Fourier
analysis of $P (\tau )$. However, the signature of the period
spacings between the different peaks with different azimuthal
orders sometimes dominates and hampers the measurement of $q$.
This most often appears on the RGB, when the rotational splittings
of the largest peaks near $\numax$ are comparable to a simple
fraction of
the frequency differences between two consecutive mixed modes.\\

Individual checks based on the fit of the mixed-mode pattern and
on \'echelle diagrams were performed on 5\,\% of the target to
correct spurious values.

\subsection{Threshold and uncertainties}

The examination of various stars at different evolutionary stages
allowed us to define a relevant threshold value $Q\ge\Qlim$
characterizing a reliable measurement of $q$. Since we noticed
that very low or very high values of $q$ are artefacts, we
introduced a penalty function for those values. Uncertainties were
empirically derived from the examination of the power spectrum $Q$
of the stretched spectrum $P ( \tau )$ and by comparison with
individual fittings. Variations of $Q$ less than 4\,\% of the
maximum value were used to derive an estimate of the uncertainties
$\dq$ (Fig.~\ref{fig-exemple}). This is a conservative value.

We are aware that a more sophisticated statistical analysis is
desirable for deriving stronger estimates of the uncertainties on
$q$, as recently done by \cite{2016A&A...588A..82B}. They chose
three bright stars seen pole-on, so with a high signal-to-noise
ratio oscillation pattern free of any rotational splitting.
Despite these favorable conditions, their analysis was
computationally very demanding (Buysschaert, personal
communication). An analysis aiming at measuring precise
uncertainties for a large set of stars is beyond the scope of our
work, which is mainly intended to provide a coherent view on a
large set of stars.

\section{Results\label{observations}}

\subsection{Data set}

We used the data of \cite{2016A&A...588A..87V}, namely a catalog
of about 6\,100 red giants with measured period spacings and duly
identified evolutionary stages. We also considered 33 subgiants,
defined as subgiants according to the seismic criterion $(\Dnu /
36.5\,\mu\hbox{Hz})(\Tg / 126\,\hbox{s}) > 1$ introduced by
\cite{2014A&A...572L...5M}. This latter work also provided us with
the criteria necessary to identify the other evolutionary stages
(RGB and clump stars). Stellar masses were estimated with the
method of \cite{2013A&A...550A.126M}, which uses the homology of
red giant oscillation spectra to lower the uncertainties induced
by pressure glitches that are present at all evolutionary stages
\citep{2015A&A...579A..84V}. This procedure benefits from a
calibration on a large set of stars and has proved to be less
biased than similar methods calibrated on the Sun only. The
calibration does not depend on the evolutionary stage, which can
lower its precision \citep{2012MNRAS.419.2077M}. Recent studies
all converge to state that seismic masses are slightly
overestimated by about 5 -- 15\,\%
\citep[e.g.,][]{2014ApJ...785L..28E,2015A&A...580A.141L,2016ApJ...832..121G};
such a result does not invalidate the relevance of the seismic
estimate.

\subsection{Iterations for the RGB and the red clump}

An iteration process allowed us to correct values of $q$ affected
by an initial incorrect estimate of $\Tg$. Finally, we obtained
$\nombreq$ values of $q$, corresponding to a maximum $Q$ value
larger than $\Qlim$. High-quality values characterized by $Q \ge
50$ were obtained for about 3700 stars. Results are shown in
Fig.~\ref{fig-coupling}, where the RGB and clump stars can be
easily identified since they show different variations with
stellar evolution. The mass dependence visible in
Fig.~\ref{fig-coupling} is a consequence of the evolution
dependence, so that its study requires some care (see Section
\ref{discussion}).

The mean value of $q$ on the RGB decreases with stellar evolution
from 0.18 to 0.12. Coupling factors in the red clump are most
often in the range [0.2 -- 0.45], with a mean value of about 0.32;
more than 70\,\% of the red-clump values are above 1/4. This
indicates that weak coupling (Eq.~\ref{eqt-couplage-Unno}) does
not hold at this evolutionary stage. For interpreting observed
coupling factors of clump stars, a theoretical study of strong
coupling must be considered \citep{2016PASJ...68..109T}.

When the threshold level of the detection is increased, the spread
in $q$ decreases for clump stars. We observed that low $Q$ values
are often associated to non-evenly spaced period spectra,
regardless of the signal-to-noise ratio of photometric time
series. As a result, low or high $q$ values are most often related
to buoyancy glitches, as observed in \cite{2015A&A...584A..50M}.

\begin{figure}
\includegraphics[width=9.1cm]{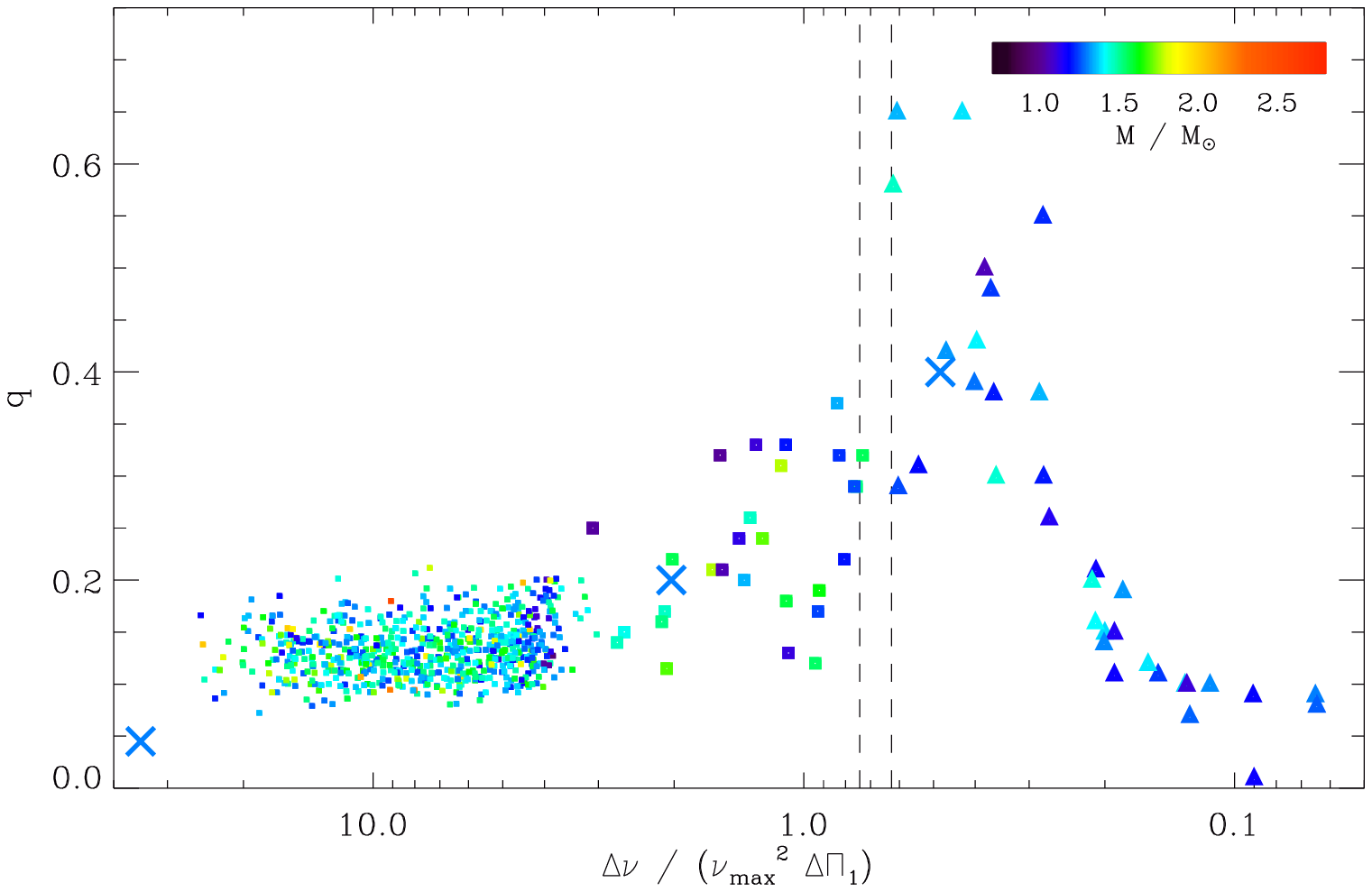}
\caption{Coupling factor $q$ as a function of the mixed-mode
density $\dens = \Dnu /(\Tg\ \numax^2)$. The vertical dashed lines
indicate the transition from subgiants  to red giants defined by
\cite{2014A&A...572L...5M}. Triangles stand for subgiants; squares
for stars on the RGB (small squares for long-cadence \Kepler\
data, big squares for short-cadence data). Three values of $q$
obtained for a synthetic 1.3-$M_\odot$ evolutionary sequence are
also shown with $\large\times$ (see Section \ref{modelling} and
Table \ref{tab-modeles}). }\label{fig-coupling-sub}
\end{figure}

\subsection{Subgiants}

Subgiants were also considered. In that case, coupling factors
were not measured with the aforementioned method, but directly
determined from the fit of the mixed-mode pattern. The validity of
the asymptotic expression is questionable for mixed modes with
low-radial gravity orders observed in red giants. It however
provides period spacings that fully agree with modelled values
\citep{2013ApJ...767..158B,2014ApJ...781L..29B,2014A&A...564A..27D}.
In fact, even if the density of gravity modes expressed by the
number $ \dens = \Dnu / (\numax^2 \Tg)$ is small, the distribution
of the period spacings is well reproduced by the asymptotic
expression and is sensitive to $q$. From the quality of the fit of
the mixed mode, we could measure this parameter and estimate
relative uncertainties smaller than 20\,\%. The identification of
high $q$ values is especially clear: in such cases, the period
spacings hardly depend on the nature of the mixed mode
(Fig.~\ref{fig-vari-q}). In Fig.~\ref{fig-coupling-sub}, $q$ is
plotted as a function of the mixed-mode density $\dens$ instead of
$\Dnu$, in order to emphasize the change of physics when a
subgiant evolves into a red giant. The strongest values of $q$,
close to unity, are obtained at the transition from subgiants to
red giants.

\begin{figure}
\begin{center}
\includegraphics[width=5.8cm]{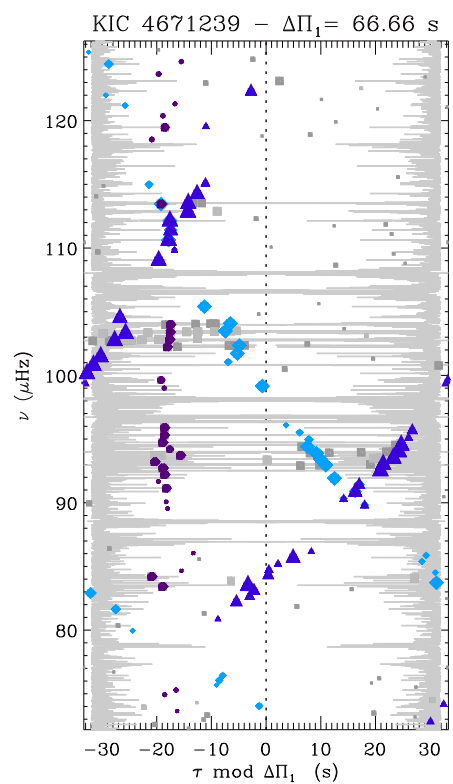}
\end{center}
\caption{\'Echelle diagram of the very low metallicity RGB star
KIC 4671239, with the stretched period $\tau$ modulo $\Tg$ on the
x-axis and frequency on the y-axis. All dark symbols are dipole
modes with a height above 5 times the local stellar background;
their sizes depend on the mode height. When the azimuthal order
can be automatically identified,  triangles correspond to $m=-1$,
squares to $m=0$,  diamonds to $m=+1$. Light grey symbols
correspond to unidentified modes: most of them are located in the
region of pressure-dominated mixed modes; a few of them may also
be $\ell=3$ modes.  All components of the rotational multiplets
intersect at 112\,$\mu$Hz; the $m=\pm 1$ components intersect at
91\,$\mu$Hz. The oscillation spectrum is plotted in gray top to
bottom in the background of the figure in order to link the
stretched periods and the mixed mode
frequency.}\label{fig-ech-hennes}
\end{figure}

\subsection{Outliers}

A limited number of stars have coupling factors significantly
different from the mean behavior. We stress that their
identification in this work dedicated to ensemble asteroseismology
is certainly incomplete; outliers are however rare. The ones we
have identified deserve future care.

\subsubsection{KIC 4671239}

KIC 4671239 has been identified by \cite{2012A&A...543A.160T} as
one of the less metallic red giants observed by \Kepler , with
[Fe/H]$ = -2.45$ and $\Teff = 4900$\,K. Its seismic parameters are
highly atypical. Apart from the high coupling value for an RGB
star at a similar evolutionary stage ($q\simeq 0.25$ instead of
less than 0.15), the analysis of its mixed-mode pattern has
revealed a period spacing $\Tg \simeq 66.6$\,s,  much smaller than
for RGB stars with similar $\Dnu$ \citep{2014A&A...572L...5M,
2016A&A...588A..87V}, and a much higher core rotation
\citep{2012A&A...548A..10M} with $\dnurot \simeq 830$\,nHz. These
parameters translate in a complex mixed-mode pattern where the
different azimuthal orders show intricate structures.
Disentangling this pattern is possible only with an \'echelle
diagram constructed with stretched periods
(Fig.~\ref{fig-ech-hennes}), following \cite{2015A&A...584A..50M}
and \cite{gehan}.

\subsubsection{KIC 6975038}

KIC 6975038 also shows atypical seismic parameters, with
$\Tg\simeq 57.9$\,s, much smaller than comparable stars with
$\Dnu=10.61\,\mu$Hz, and $q \simeq 0.35$, much above the values
observed on the RGB. We note that the low value of $\Tg$ and the
high value of $q$ may both indicate a larger radiative cavity,
hence a smaller region where modes are evanescent, than in other
stars with similar $\Dnu$.

\subsubsection{Massive stars}

Stars in the secondary clump, more massive than 1.8\,$M_\odot$,
show lower $q$ compared to less massive stars burning their
core-helium. The difficulty to fit their mixed-mode spectrum does
not resemble to the difficulty met in presence of buoyancy
glitches \citep{2015A&A...584A..50M}. A better fit of the data can
be obtained with a gradient in $q$ as a function of frequency.
According to the study conducted in Section \ref{formalism}, this
gradient may result from the fact that the $\NBV$ and $S_1$
profiles are not parallel in the evanescent region of such stars.

This effect appears for stars as KIC 4372082 or 6878041, more
massive than the secondary-clump stars studied by
\cite{2015A&A...580A..96D}. Such objects are rare and, despite
low-quality spectra, certainly deserve detailed study and
modelling.

\section{Discussion\label{discussion}}

The variation of the coupling factor with the large separation
$\Dnu$ can be used to derive direct information on the extent of
the evanescent region between the $\NBV$ and $S_1$ profiles, with
the formalism introduced in Section \ref{formalism}. Even if mixed
modes probe these functions in a limited frequency range around
$\numax$, the hypothesis that the $\NBV$ and $S_1$ profiles are
parallel in the considered region helps us in deriving precise
information on stellar interior properties, regardless of the
consequence of the decrease of $\numax$ with evolution. The
hypothesis is finally discussed in Section~\ref{section-hyp}.

\subsection{Variation with the evolutionary stage}

\subsubsection{Subgiant - red giant transition}

\cite{2014A&A...572L...5M} have shown that the transition between
subgiants and redgiants has a clear signature in the $\Tg$ --
$\Dnu$ diagram. In the subgiant phase, the relation between these
asymptotic period and frequency spacings shows a large spread with
a significant mass dependence, whereas the parameters are tightly
bound on the RGB. The coupling factor $q$ is similarly impacted by
the evolution and shows a significant increase at the end of the
subgiant phase (Fig.~\ref{fig-coupling-sub}). This increase occurs
in parallel with the first dredge-up, when the base of the
convection zone goes deeper in the envelope. A large value of $q$
indicates that not only the $\NBV$ profile but also the $S_1$
profile is deep in the stellar interior. Observing high values in
the range [0.40 -- 0.65] near the transition region from subgiants
to red giants argues, using Fig.~\ref{fig-qNS}, for a slope
$\beta$ slightly higher than or close to 1.3 with a ratio
$\alpha^2$ larger than about 0.5.


\begin{table}
\caption{Mass dependence of $q$ on the RGB}\label{tab-RGB-masse}
\begin{tabular}{ccccccc}
\hline
 $M/M_\odot$& 1.0   & 1.2   & 1.4   & 1.6   & 1.8   & 2.0  \\
 \hline
  $q_M$     & 0.144 & 0.128 & 0.126 & 0.123 & 0.134 & 0.134\\
$\sigma_q$  & 0.021 & 0.025 & 0.026 & 0.022 & 0.023 & 0.027\\
 \hline
\end{tabular}

The variation of $q$ with $\Dnu$ is modelled as $q = q_M (\Dnu /
10)^{0.096}$, with $\Dnu$ in $\mu$Hz.
\end{table}

\begin{table}
\caption{Mass dependence of $q$ in the red
clump}\label{tab-clump-masse}
\begin{tabular}{ccccccc}
\hline
 $M/M_\odot$ &  0.9  &  1.1  &  1.3  &  1.5  &  1.7  &   2.0   \\
\hline
 $q\ind{b}$  &       & 0.301 & 0.287 & 0.274 & 0.272 &   \\
 $\bar{q}  $ & 0.363 & 0.324 & 0.299 & 0.289 & 0.283 & 0.270\\
 $q\ind{e}$  &       & 0.339 & 0.313 & 0.303 & 0.290 &  \\
 $\sigma_q$  & 0.024 & 0.016 & 0.011 & 0.010 & 0.012 & 0.022\\
\hline
\end{tabular}

$q\ind{b}$, $\bar{q}$ and $q\ind{e}$ measure the coupling for
respectively the early, middle and late stages in the clump,
defined in \cite{2014A&A...572L...5M}; these values are defined by
25\,\% of the stars lying on the first, middle, and late portion
of the mass-dependent evolutionary track, respectively; $\sigma_q$
measures the spread of the values; the fits were derived from
high-quality spectra with $Q\ge 50$.
\end{table}

\subsubsection{On the RGB}

At the beginning of the ascent on the RGB, the stellar core
contracts and the envelope expands. The decrease of $q$ means
either that the region between $\NBV$ and $S_1$ expands too,
inducing a decrease of $\alpha$, or that the coefficient $\beta$
increases toward the value 3/2. A simultaneous variation of both
terms $\alpha$ and $\beta$ is possible too.

For evolved models on the RGB, $\numax$ becomes smaller than the
value of $\NBV$ at the base of the convective zone, so that the
hypothesis of parallel variations of $\NBV$ and $S_1$ is no longer
valid for evolved stars on the RGB
\citep[e.g.,][]{2013ApJ...766..118M}. In fact, the decrease of $q$
with stellar evolution can result either from the decrease of
$\numax$ or from the expansion of the region between $\NBV$ and
$S_1$, so that it is impossible to derive any firm conclusion in
terms of interior structure evolution.

The decrease of $q$ on the RGB plays a non-negligible role in the
difficulty to observe mixed modes at small $\Dnu$ predicted for
high- and low-mass stars
\citep{2009A&A...506...57D,2014A&A...572A..11G}. In fact, mixed
modes that are not in the close vicinity of the pressure-dominated
modes are poorly coupled, so that they show an important gravity
character, hence a high inertia and a tiny amplitude. In practice,
measuring low values of $q$ is hard.

In order to estimate the mass dependence of $q$ observed on the
RGB, we first fitted the slope of $q(\Dnu)$ as a power law, under
the assumption that the exponent does not depend on the stellar
mass. In a second step, we quantified the mass dependence in the
relation $q(\Dnu)$. For low-mass stars with a degenerate helium
core, the higher $M$, the lower $q$; for stars more massive than
1.8\,$M_\odot$, the situation is inverted
(Table~\ref{tab-RGB-masse}). This change occurs near the limit in
mass between the red and secondary clumps, so is likely related to
the degeneracy of helium in the  core.

\subsubsection{In the red clump}

In the red clump, the mass dependance of $q$ is coupled to the
large separation dependence: the lower the mass, the higher the
coupling. This behavior is discussed in the next paragraph, since
it obeys to a general trend in the relationship between $q$ and
$\Tg$.

Mean values of the coupling for red clump stars are given in Table
\ref{tab-clump-masse}. Using the seismic evolutionary tracks
depicted in \cite{2014A&A...572L...5M}, we could measure the
evolution of $q$ in the red clump, at fixed mass. Values at the
early, middle and late stages in the clump are shown in Table
\ref{tab-clump-masse}.  As shown by \cite{2015A&A...584A..50M},
the evolution of stars in the red clump show non-monotonous
variation of both $\Dnu$ and $\Tg$. Conversely, the monotonous
increase of $q$ indicates either that the ratio $\alpha$ increases
along the evolution of low-mass stars in the red clump, resulting
from a small shrinking of the evanescent region, or that the
exponent $\beta$ decreases. Both effects may simultaneously
contribute to the variations.

\subsection{Variation with the period spacing}

\begin{figure}
\includegraphics[width=8.99cm]{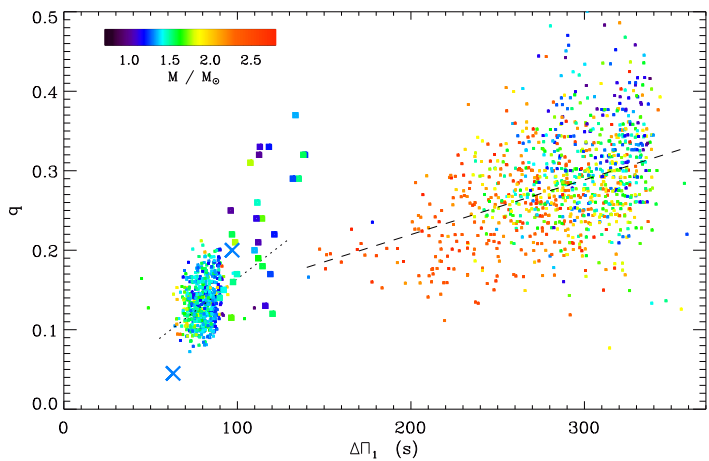}
\caption{Coupling factor $q$ as a function of the period spacing
$\Tg$ for RGB and clump stars. Same symbols as in
Fig.~\ref{fig-coupling-sub}. A threshold at $Q=100$ reduces the
number of stars compared to Fig.~\ref{fig-coupling}. The dotted
and dashed lines correspond to the fits for RGB and clump stars,
respectively. }\label{fig-coupling-DPi1}
\end{figure}

The variations of $q$ with the period spacing $\Tg$ depend on the
evolutionary stage. Either on the RGB or in the clump, the global
variations of $q(\Tg)$ indicate that the larger $\Tg$, the larger
$q$ (Fig.~\ref{fig-coupling-DPi1}). A large value of $\Tg$ is
representative of a small dense core with high $\NBV$ values. So,
we come to the conclusion that high values of $\NBV$ and $S_1$
occur in similar situations, and that they get close to each other
when they increase together. At fixed $\beta$, the ratio $\alpha =
\NBV / S_1$ is then correlated with $\Tg$ for both RGB and clump
stars (but not for subgiants).

Since the variations of $q$ are nearly linear in both domains,
simple fits can be used as proxies for $q$
(Fig.~\ref{fig-coupling-DPi1}):
\begin{eqnarray}
  q\ind{RGB}  &=& -0.0034 + {\Tg \over 597}, \\
  q\ind{clump}&=& 0.082   + {\Tg \over 1450},
\end{eqnarray}
where $\Tg$ is measured in seconds. The fit of $q\ind{RGB}$ is
however not efficient for early RGB stars; the fit of
$q\ind{clump}$ is valid for both primary and secondary clumps.
Such fits are intended to facilitate the identification of the
mixed-mode pattern, not for explaining the physics of the
coupling.

\begin{table}
\caption{Seismic properties of 1.3\,$M_\odot$
models}\label{tab-modeles}
\begin{tabular}{ccccccc}
\hline
  Model &Evol.& Age  & $\numax$ & $\Dnu$   & $\Tg$& $q$ \\
        &stage       & (Gyr)& ($\mu$Hz)& ($\mu$Hz)& (s)  &    \\
 \hline
    M0  & subgiant   & 4.40 & 670      & 39.5     & 182  &  0.40 \\
    M1  & RGB        & 4.59 & 341      & 22.9     & 97   &  0.20 \\
    M2  & RGB        & 4.79 & 49.0     & 5.24     & 63   &  0.045 \\
\hline
\end{tabular}
\end{table}

\subsection{Modelling and computation of $q$\label{modelling}}

We used stellar models to match the observed coupling factor. We
considered 1.3-$M_\odot$ models at three evolutionary stages
(Table~\ref{tab-modeles}), without overshoot or diffusion; their
full description is given in \cite{2015A&A...579A..31B}. In the
strong coupling case (model M0), we computed $q$ from
Eqs.~(\ref{eqt-relation-T-q})--(\ref{eqt-definition-X}) using the
asymptotic formalism of \cite{2016PASJ...68..109T} and taking the
perturbation of the gravitational potential into account. In the
intermediate case (model M1), using a similar analysis is
questionable since hypotheses in the calculation of $\XR$ are
valid when the evanescent region is very thin. Nevertheless, the
strong-coupling analysis matches the weak-coupling case when the
evanescent region becomes very large, so that we computed $q$ in a
similar way as for the model M0. In the weak coupling case (model
M2), we used the same expression, but with $\XR = 0$, and finally
compare these three factors with observations. They show in fact
small variations with frequency, so that we had to consider their
mean value, defined in a 4-$\Dnu$ broad frequency range centered
on $\numax$. We checked that the term $\XR$ introduced by
Eq.~(\ref{eqt-definition-X}) significantly reduces the value of
$q$ and also contributes to the relative stability of $q$ with
frequency. We also noticed that the perturbation of the
gravitational potential plays a non-negligible role not only in
the core but also in the evanescent zone and in the inner region
of the convective envelope. This provides evidence that the
Cowling approximation is not appropriate for computing $q$.

Modelling quantitatively agrees with observations, except for the
most evolved model. The subgiant model M0 close to the transition
to red giants shows a high $q$; the next model M1 is on the RGB
and has a much lower $q$. Both  agree with the observed values.
The coupling factor for model M2, higher on the RGB, shows however
a value significantly smaller than observed. In M2, the evanescent
region is in fact above the base of the convective envelope
\citep[e.g., Fig.~2 of][]{2013ApJ...766..118M}. This point
deserves further work, beyond the scope of this paper.

Modelling red clump stars requires special care in the
prescription of convection and mixing in the core
\citep{2012A&A...543A.108L,2015MNRAS.453.2290B}. We could not
compare our results with modelling but used the calculations
exposed in Section \ref{formalism}. The high values observed for
$q$ in the clump imposes the exponent $\beta$
(Eq.~\ref{eqt-def-beta}) to be less than $3/2$
(Fig.~\ref{fig-qNS}).

\subsection{Variation of $q$ with frequency\label{section-hyp}}

For a limited number of stars examined by
\cite{2016A&A...588A..87V}, the asymptotic expansion does not
provide a satisfying fit of the mixed-mode pattern, without a
direct explanation in terms of buoyancy glitch. For these stars, a
better fit is obtained when varying $q$ with frequency. Modelling
derives similar conclusion, with different slopes $\beta_N$ and
$\beta_S$ (Eq.~\ref{eqt-def-beta}) or more complicate variations
\citep[e.g.,][]{2013ApJ...766..118M}.

Since the independence of $q$ with frequency derives from the
hypothesis of parallel variations of $\NBV$ and $S_1$, we have to
conclude that this hypothesis is not fully correct. The
observational study of the variations of $q$ with frequency
appears to be highly challenging, for the same reasons as those
explaining the difficulties in measuring $q$ (rotation, glitches
and finite mode lifetimes). The individual study of bright stars
with a high signal-to-noise ratio is necessary to investigate the
frequency dependence of $q$ in detail. Combined with modelling,
this study will help assessing to which extent the mean value of
$q$ is  a global seismic parameter as informative for the
evanescent region as $\Dnu$ and $\Tg$ for the pressure and
radiative cavities, respectively.

\section{Conclusion\label{conclusion}}

The new method setup for measuring the coupling factor of mixed
modes in evolved stars has provided the first analysis of this
parameter over a large set of stars. We could determine {\nombreq}
values of $q$, from subgiants to clump stars. Three main results
can be inferred:

- Coupling factors test the region between the \BV\ cavity and the
$S_1$ profile, dominated by the radiative core and the
hydrogen-burning shell. Indeed, while $\Dnu$ is mainly sensitive
to the envelope and $\Tg$ provides the signature for the core, we
can directly access the intermediate region with $q$.

- The variation of $q$ with stellar evolution provides us with new
constraints on stellar modelling. We note that the coupling
factors show simple global variations with the period spacing, for
both RGB and clump stars. The characterization of outliers can be
used to constrain physical processes inside stars.

- Strong coupling is observed in stars at the transition
subgiant/red giant and in the red clump. In fact, the
weak-coupling formalism of \cite{1989nos..book.....U} fails for a
quantitative use of the observed coupling factors, except for the
most evolved stars on the RGB. In all other cases, the use of the
new formalism proposed  by  \cite{2016PASJ...68..109T} for strong coupling is mandatory.\\

These measurements also open the way to more precise fits of the
mixed-mode pattern, in order to analyze in detail extra features,
as the rotational splittings and the buoyancy glitches.

\begin{acknowledgements}

We acknowledge the entire \emph{Kepler} team, whose efforts made
these results possible.  BM, CP, and KB acknowledge financial
support from the Programme National de Physique Stellaire
(CNRS/INSU), from the French space agency CNES, and from the ANR
program IDEE Interaction Des \'Etoiles et des Exoplan\`etes. MT is
partially supported by JSPS KAKENHI Grant Number 26400219. MV
acknowledges funding by Funda\c c\~ ao para a Ci\^encia e a
Tecnologia (FCT) through the grant CIAAUP-03/2016-BPD, in the
context of the project UID/FIS/04434/2013, co-funded by FEDER
through the program COMPETE2020 (POCI-01-0145-FEDER-007672).

\end{acknowledgements}
\bibliographystyle{aa} 

\end{document}